\documentclass[runningheads]{llncs}
\usepackage{booktabs}
\usepackage{graphicx}
\usepackage{array}
\newcolumntype{L}[1]{>{\raggedright\let\newline\\\arraybackslash\hspace{0pt}}m{#1}}
\newcolumntype{C}[1]{>{\centering\let\newline\\\arraybackslash\hspace{0pt}}m{#1}}
\newcolumntype{R}[1]{>{\raggedleft\let\newline\\\arraybackslash\hspace{0pt}}m{#1}}
\usepackage{xcolor}
\usepackage{hanging}
\usepackage{lipsum}
\usepackage{caption}
\usepackage{subcaption}
\usepackage{cite}
\usepackage{amsmath}

\newenvironment{RQquestion}{%
  \par%
  \leftskip=1em\rightskip=1em%
  \noindent
  \ignorespaces}
  
\newenvironment{RQanswer}{%
  \par%
  \leftskip=3em\rightskip=2em%
  \noindent
  \ignorespaces}{%
  \par\medskip}

\begin{document}
%
\title{Predicting the Number of Reported Bugs in a Software Repository}
%
%
\author{Hadi Jahanshahi \and
Mucahit Cevik \and
Ay\c{s}e Ba\c{s}ar}
\authorrunning{H. Jahanshahi et al.}

\institute{Data Science Lab at Ryerson University, Toronto, ON M5B 1G3, Canada
\email{\{hadi.jahanshahi,mcevik,ayse.bener\}@ryerson.ca}}

\maketitle

\begin{abstract}
The bug growth pattern prediction is a complicated, unrelieved task, which needs considerable attention. Advance knowledge of the likely number of bugs discovered in the software system helps software developers in designating sufficient resources at a convenient time. The developers may also use such information to take necessary actions to increase the quality of the system and in turn customer satisfaction. In this study, we examine eight different time series forecasting models, including Long Short Term Memory Neural Networks (LSTM), auto-regressive integrated moving average (ARIMA), and Random Forest Regressor. Further, we assess the impact of exogenous variables such as software release dates by incorporating those into the prediction models. We analyze the quality of long-term prediction for each model based on different performance metrics. The assessment is conducted on Mozilla, which is a large open-source software application. The dataset is originally mined from Bugzilla and contains the number of bugs for the project between Jan 2010 and Dec 2019. Our numerical analysis provides insights on evaluating the trends in a bug repository. We observe that LSTM is effective when considering long-run predictions whereas Random Forest Regressor enriched by exogenous variables performs better for predicting the number of bugs in the short term.  

\end{abstract}
\section{Introduction}
Bug prediction is a crucial task in software engineering practice as it lends insight for practitioners to prepare their resources before their system becomes overwhelmed with defects. Predicting the number of reported bugs to a software system enables developers, managers, and product owners to distribute limited resources, take timely decisions towards effort reduction, and maintain a high-level of software quality. Therefore, there is a need for an automated bug number estimation to facilitate decision making in software development. 

The bug growth pattern is a complex and tedious task, and there is uncertainty in the reporting time, assigning time, and fixing time of a bug~\cite{Herraiz2007}. Despite the random bug introduction pattern, there are certain rules and patterns in those interactions which can be a valuable source of information~\cite{Pati2018}. In this paper, we have extracted the number of bugs from the Mozilla bug repository from Jan 2010 to Dec 2019. The data are split to the weekly bug number, and there exist 522 weeks in total. As an exogenous factor, we extracted the release times of Mozilla updates to see whether multivariate modelling can enhance the performance of the prediction models. 

Previous studies on modelling the bug growth patterns mostly use generic time series models, e.g. auto-regressive integrated moving average (ARIMA), X12 enhanced ARIMA, exponential smoothing, and polynomial regression~\cite{Wu2010,Rastogi2014,Kenmei2008,Shariat2016,Krishna2018}.We note that the previous studies lack a rational baseline to compare our methods against. In most cases, they compared different algorithms with each other without having a concrete baseline. To alleviate the issue, we defined lag(1) prediction as a naive baseline - that is, predicted value for the target step is exactly equal to the last observed value. The worst-case scenario is that the prediction model should outperform the naive baseline, otherwise the model can be considered useless. 

Time series prediction models have been used in the software engineering domain for the past 20 years. Chora{\'{s}} et al.~\cite{choras2019} exploit time series methods such as ARIMA, Holt-Winters, and random walk to forecast various project-related characteristics. Pati and Shukla~\cite{Pati2014} compared ARIMA and Artificial Neural Network models for Debian Bug Number Prediction. Their comparative analysis indicates that the combination of ANN and ARIMA improves prediction accuracy. Destefanis et al.~\cite{Giuseppe2017} used time series analysis to determine seasonality and trends of Affective Metrics for Software Development. They consider the evolution of human aspects in software engineering. Our study is different than these studies in that we focused on the reported number of bugs in software as a metric which helps developers maintain the software quality.   

Wang and Zhang~\cite{Wang2012} use a different approach to predict Defect Numbers. They design Defect State Transition models and apply the Markovian method to predict the number of defects at each state in the future. There are also studies that consider software defect number prediction in method-level and file-level~\cite{CHEN2019161,Graves2000,Gao2007}. Our work differs from these two studies as we consider the bug reported to the system regardless of whether it is valid or not rather than the number of bugs existing in different granularities of the system. Hence, we set out to study different approaches that are not well investigated to identify the number of bugs reported to the Mozilla project. We structure our study along with the following two research questions:

\begin{RQquestion}
    \textbf{RQ1: How accurately the number of bugs in a project can be predicted using time series analysis?}
\end{RQquestion}
\begin{RQanswer}
  Time series prediction assumes that there are some patterns in the time series, making it feasible for prediction. Nonetheless, it is not always the case. We tried multiple time series models, including Long Short Term Memory (LSTM), ARIMA, Exponential Smoothing (EXP), Weighted Moving Average (WMA), and Random Forest (RF) regressor with or without exogenous features. Surprisingly, the performance of a one-step prediction for all models is not significantly different. Furthermore, the baseline seems as good as the others, a new finding which was not considered in previous studies. 
\end{RQanswer}

\begin{RQquestion}
    \textbf{RQ2: How feasible is long-term bug number prediction?}
\end{RQquestion}
\begin{RQanswer}
  Not all models are able to predict more than one step ahead. Hence, we investigate the feasibility and sensitivity of different models to long-term prediction. Specifically, we consider a 3-month prediction (or equivalently 13-week prediction). For the Mozilla project, LSTM shows a significant improvement compared to traditional time series models. The performance of the model is almost of the same quality as it was for a one-step prediction.
\end{RQanswer}

The rest of the paper is organised as follows. Section~\ref{methodolgy} discusses the experimental setting of the models, including the preprocessing phase, the datasets, and a brief discussion of forecasting models. Section~\ref{sec:res} presents the performance results of different models for predictions over a long and short horizon. Section~\ref{sec:threats} discloses the threats to the validity of our findings. Finally, Section \ref{sec:conclusion} concludes the paper. 

\section{Methodology} \label{methodolgy}
To predict the number of introduced bugs in the system, we used both statistical models and machine learning techniques. Before applying time series models, we first check certain requirements to ensure that data is stationary~\cite{jothimani2019stock}. We conduct Augmented Dickey-Fuller (ADF) for this purpose. The ADF test determines the number of lags by the Akaike information criterion (AIC). The null hypothesis (at the significance level $\alpha = 0.05$) of the test is that the data are non-stationary~\cite{Dicky-fuller}. The $p$-value of the test is $0.012$ thus rejecting the null hypothesis. This suggests that the time series does not have a unit root, and in turn it is stationary. Hence, the data can be modelled directly and there is no need to have supplementary preprocessing or transformation. To further investigate the result of the test, we conduct the auto-correlation function and partial auto-correlation function as well (see Section~\ref{ACFPACF}).

\begin{figure*}[ht!]
    \centering
    \includegraphics[width=0.85\textwidth]{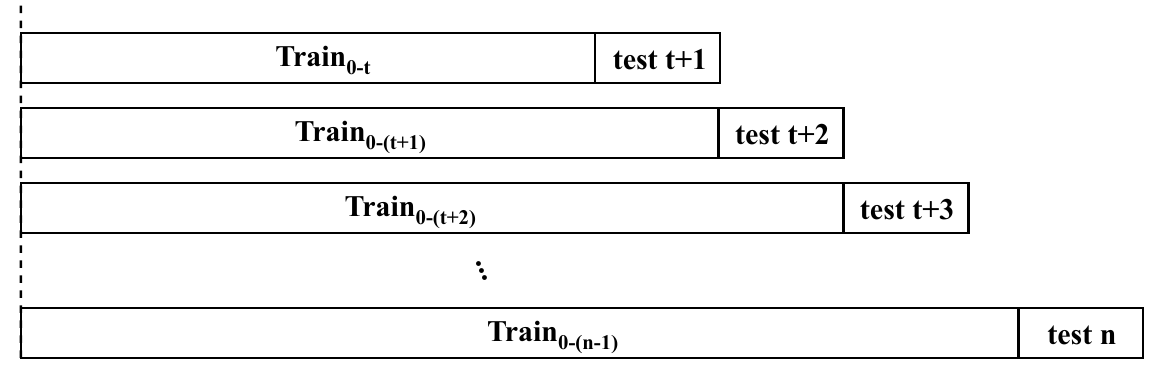}
    \caption{Train-test rolling visualisation}
    \label{fig:rolling}
\end{figure*}

In order to train time series models on the data, we applied a rolling method for training a time series dataset~\cite{TASHMAN2000437}. The idea is to train the dataset, $train_{(0-t)}$, including time series from time 0 to $t$, and test the model on time $t+1$, $test_{t+1}$. In the next step, the ground truth at time $t+1$ is added to the training set and tested on $test_{t+2}$. Figure~\ref{fig:rolling} demonstrates a visual representation of this approach. 

We used the weekly reported bug dataset from January 2010 to January 2017 as the training set and February 2017 to December 2019 as the test set (see Section~\ref{dataset}). After the training-test split based on the rolling approach, different models have been evaluated for the given dataset (see Section~\ref{models}).

\subsection{Data}\label{dataset}
We have extracted the number of reported bugs from the Mozilla bug repository~\footnote{Mozilla Bug Tracking System. https://bugzilla.mozilla.org/}. Mozilla is established in 1998 by Netscape as an open-source community. The bug-related information for Mozilla and its products is tracked in the Bugzilla system. We gather reported number of bugs data for Mozilla for the past decade (from January 2010 to the end of December 2019) with a total count of 100,450. We divided data into weekly bug number- that is 522 weeks in total. Mozilla suite weekly bug arrival ranges from 50 to 558.

\begin{figure*}[ht!]
    \centering
    \includegraphics[width=1\textwidth]{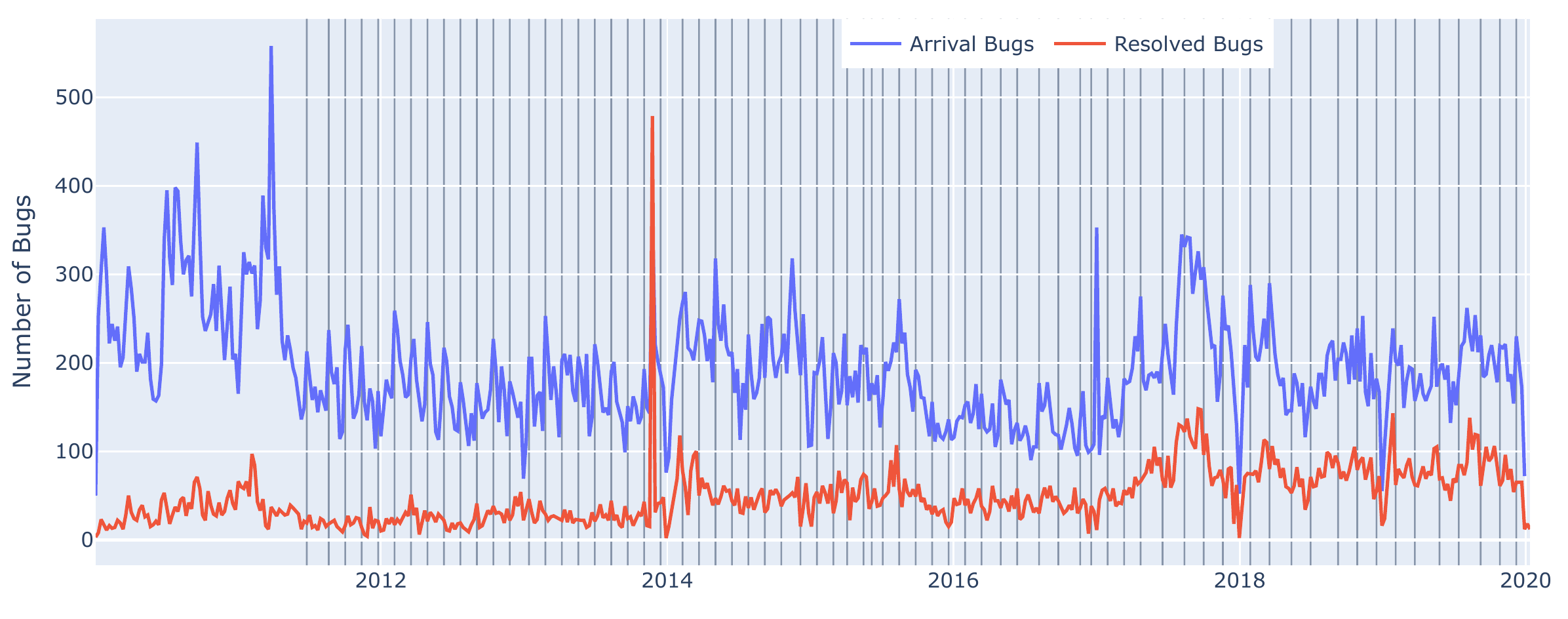}
    \caption{The number of bugs introduced and resolved in Mozilla during the past decade. Horizontal black lines indicate release dates of different versions of the Mozilla project.}
    \label{fig:bug_number}
\end{figure*}

Figure~\ref{fig:bug_number} demonstrates the arrival and resolved bugs in the Mozilla project. The clear outlier in the number of resolved bugs related to the end of 2013 is manually removed from the dataset. Here, our aim is to predict the number of bugs arriving to the system (arrival bugs) to help practitioners efficiently allocate resources. 

\subsection{ACF and PACF}\label{ACFPACF}
\begin{figure*}[ht!]
    \centering
        \begin{subfigure}[t]{0.5\textwidth}
        \centering
        \includegraphics[width=1\textwidth]{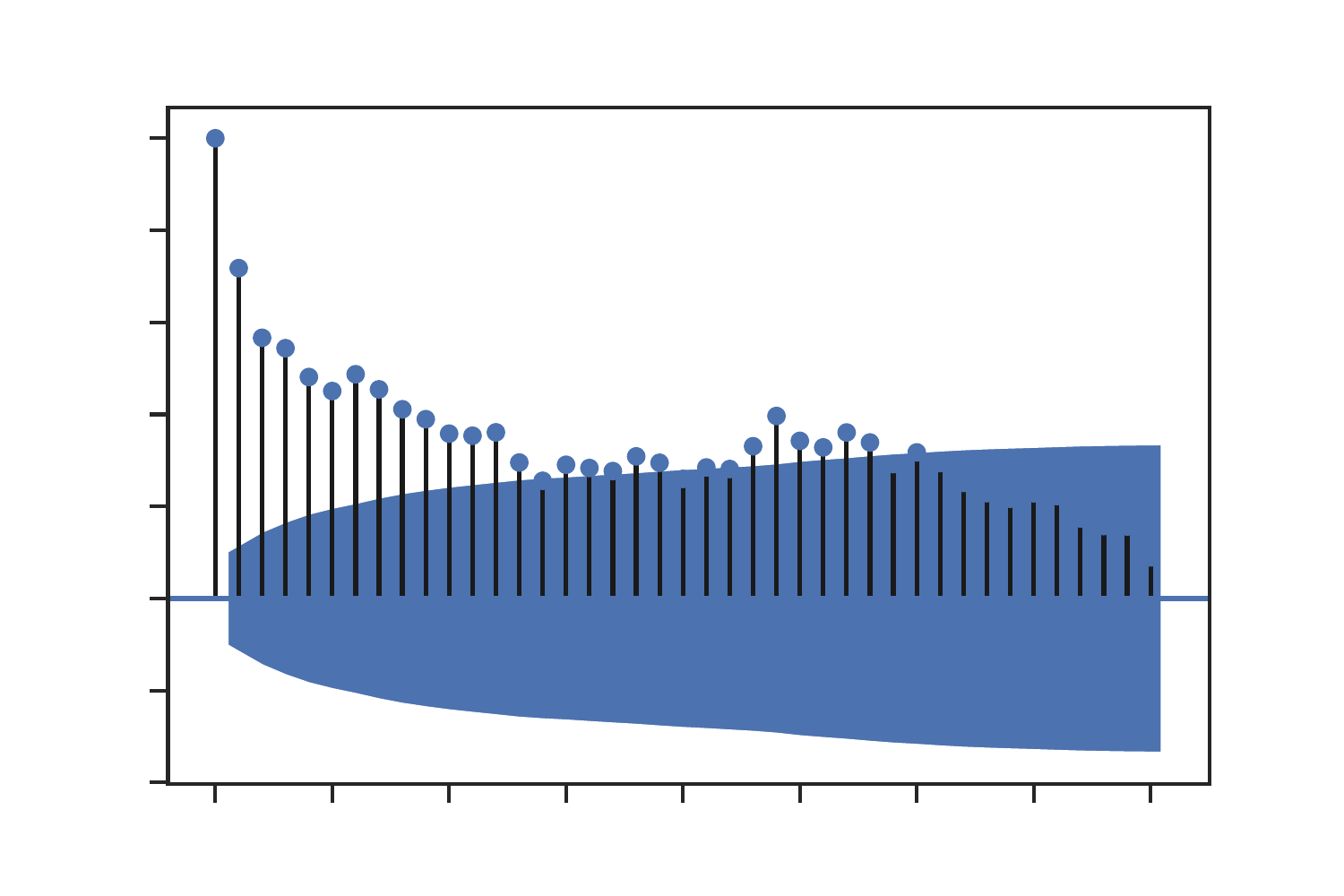}
        \vspace{-0.6cm}
        \caption{Auto-correlation Function}
        \label{fig:ACF}
    \end{subfigure}%
    ~ 
    \begin{subfigure}[t]{0.5\textwidth}
        \centering
        \includegraphics[width=1\textwidth]{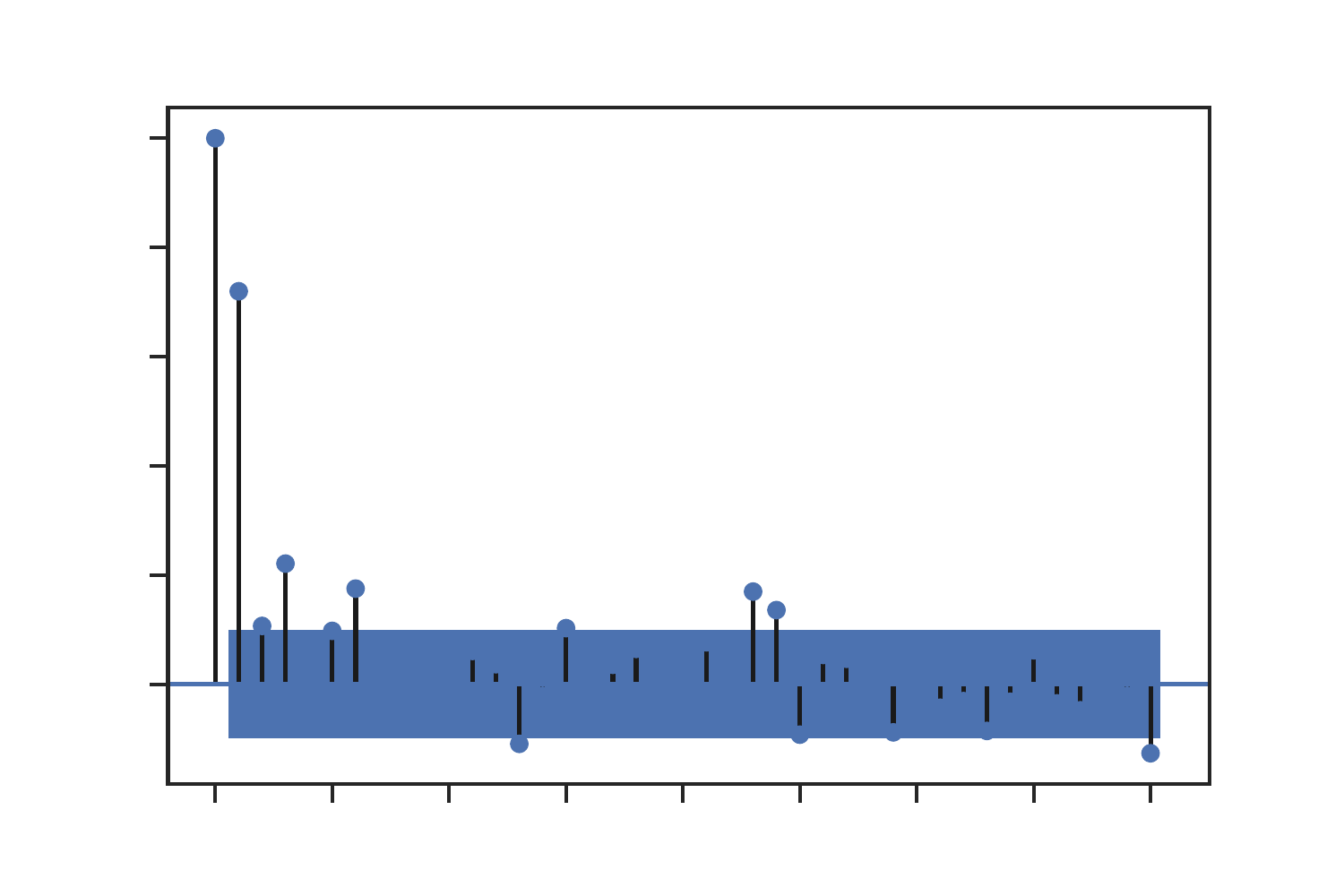}
        \vspace{-0.6cm}
        \caption{Partial Auto-correlation Function}
        \label{fig:PACF}
    \end{subfigure}
    \caption{ACF and PACF plots. ACF Long short term memory gradual decline whereas PACF cuts off sharply.}
\label{fig:ACF-PACF}
\end{figure*}

Auto-correlation function (ACF) defines the correlation of the series between different lags. In other words, it specifies how well the current value of the series is related to its past values. Therefore, this characteristic enables it to depict seasonality, cycles, and trends in data. On the other hand, partial auto-correlation function (PACF) shows the correlation of the residuals with the next lag value. The sharp cut-off in PACF represents the number of Auto-Regressive (AR) terms, $p$, while a sharp drop in ACF is associated with Moving Average (MA) terms, $q$. Figure~\ref{fig:ACF-PACF} illustrates the general tendency in ACFs and PACFs calculated on our training set. The PACF drops sharply when the lag is 2, indicating the number of AR terms, $p$, is two. ACF does not have such behaviour, so we assume $q$ is equal to zero.

\subsection{Forecasting models} \label{models}
In order to forecast the issues reported to the Mozilla project, we consider different algorithms, including LSTM, ARIMA, Exponential Smoothing, Weighted Moving Average, and RF regressor. We also defined exogenous variables (covariates) to examine the effect of external and seasonal variables on the prediction performance. The exogenous variables include \textit{Branch dates} (from version history), \textit{week of the month}, \textit{month of the year}, and \textit{year}. For the models which incorporate exogenous variables, we add ``x" at the end of their name; for instance, RFx is a random forest containing exogenous variables. The brief explanations for each model are provided below. 

\begin{enumerate}
    \item \textbf{Naive Baseline}: It assumes the number of bugs at time $t$ is equal to that at time $t-1$. Note that we expect to see a better result for the proposed models compared to naive baseline defined in this manner. 
    \item \textbf{EXP}: It considers two factors in its prediction: the forecast value at the previous timestamp and its actual value. Therefore, it is defined as
    \begin{equation*}
        X_{t} = \alpha X_{t-1} + (1-\alpha) \hat{X}_{t-1}
    \end{equation*}
    where $X_t$ and $\hat{X}_t$ are the actual and predicted values at time $t$, respectively, and $\alpha$ is the smoothing level.
    \item  \textbf{WMA}: Weighted Moving Average simply forecasts based on a weighted average of the previous steps. 
    \item \textbf{ARIMA}: It is one of the most popular models used for time series prediction~\cite{Zhang2018, choras2019, Pati2014, Pati2017}. The general ARIMA model $(p,q,d)$ is formulated as
    \begin{equation*}
        W_t = \sum_{i=1}^{p}\alpha_i W_{t-i} + \sum_{j=0}^{q}\beta_j e_{t-j}
    \end{equation*}
    where $W_{t} = X_{t} - X_{t-d}$, $\alpha_i$ and $\beta_i$ represent the linear coefficients of the model, $e(t)$ is the error concerning the mean, and $d$ is the degree of non-stationary homogeneity. The value associated with each parameter has been discussed in Section~\ref{ACFPACF}.
    \item \textbf{LSTM}: Long short-term memory is widely used for time series analysis. They capture  both long temporal dependencies and short term patterns in data. We use the LSTM cell architecture defined by \cite{LSTM} as follows:
    \setlength{\belowdisplayskip}{3pt} \setlength{\belowdisplayshortskip}{3pt}
    \begin{equation*}
        f_t=\sigma \big(W_f.[h_{t-1}, x_t]+b_f  \big)
    \end{equation*}
    \begin{equation*}
        i_t=\sigma \big(W_i.[h_{t-1}, x_t]+b_i  \big)
    \end{equation*}
    \begin{equation*}
        \hat{C}_t= \tanh\big(W_c.[h_{t-1}, x_t]+b_c  \big)
    \end{equation*}
    \begin{equation*}
        C_t= f_t \times C_{t-1} + i_t \times \hat{C}_t
    \end{equation*}
    \begin{equation*}
        o_t= \sigma \big(W_o.[h_{t-1}, x_t]+b_o  \big)
    \end{equation*}
    \begin{equation*}
        h_t= o_t \times \tanh(C_t)
    \end{equation*}
    where $b$ is the bias term in all equations. $f_t$ is the forget gate deciding which information coming from hidden state $h_{t-1}$ and the new input $x$ should be discarded. Update layer, $i_t$, selects critical information to be stored and multiplied with candidate vector, $\hat{C}_t$, generating $C_t$ vector as the result. Finally, the model decides which information should be reported, $o_t$, and which one should be passed to the next cell, $h_t$. 
    \item \textbf{RF}: Unlike the greedy nature of uni-variate trees, Random Forest is less prone to overfitting on the data as it is an amalgam of DTs which takes the random subset of features in each tree. We applied RF Regressor as a new method that has not been used in this domain. 
    
\end{enumerate}

The parameters of the models are shown in Table~\ref{tab:models_settings}. 
\setlength{\tabcolsep}{8pt}
\renewcommand{\arraystretch}{1.2}
\begin{table}[!ht]
\centering
\caption{Models' parameter settings\label{tab:models_settings}}%
\resizebox{0.72\linewidth}{!}{
\begin{tabular}{ll}
\toprule
\textbf{Model} & \textbf{Settings} \\
\midrule
EXP & $\alpha = 0.5$ \\ \hline
WMA & \begin{tabular}[c]{@{}l@{}}Number of previous steps = 2\\  Weights = {[}0.66 , 0.33{]}\end{tabular} \\
\hline
ARIMA & $[p,q,d] = [2,0,0]$ \\ \hline
LSTM & \begin{tabular}[c]{@{}l@{}}Number of units = 100\\ Number of epochs = 50\\ with log and difference transformation\end{tabular} \\ \hline
RF & \begin{tabular}[c]{@{}l@{}}Number of trees = 100\\ Number of feature = sqrt (total features)\end{tabular} \\ 
\bottomrule
\end{tabular}
}
\end{table}

\section{Results}\label{sec:res}
After designing the experiment, Root Mean Square Error (RMSE), R-squared ($R^2$), Error percentage, Median Absolute Error (MAE), and Error Stand Deviation (Std) are used to contrast the algorithms' performance. 

\begin{figure*}[ht!]
    \centering
    \includegraphics[width=\textwidth]{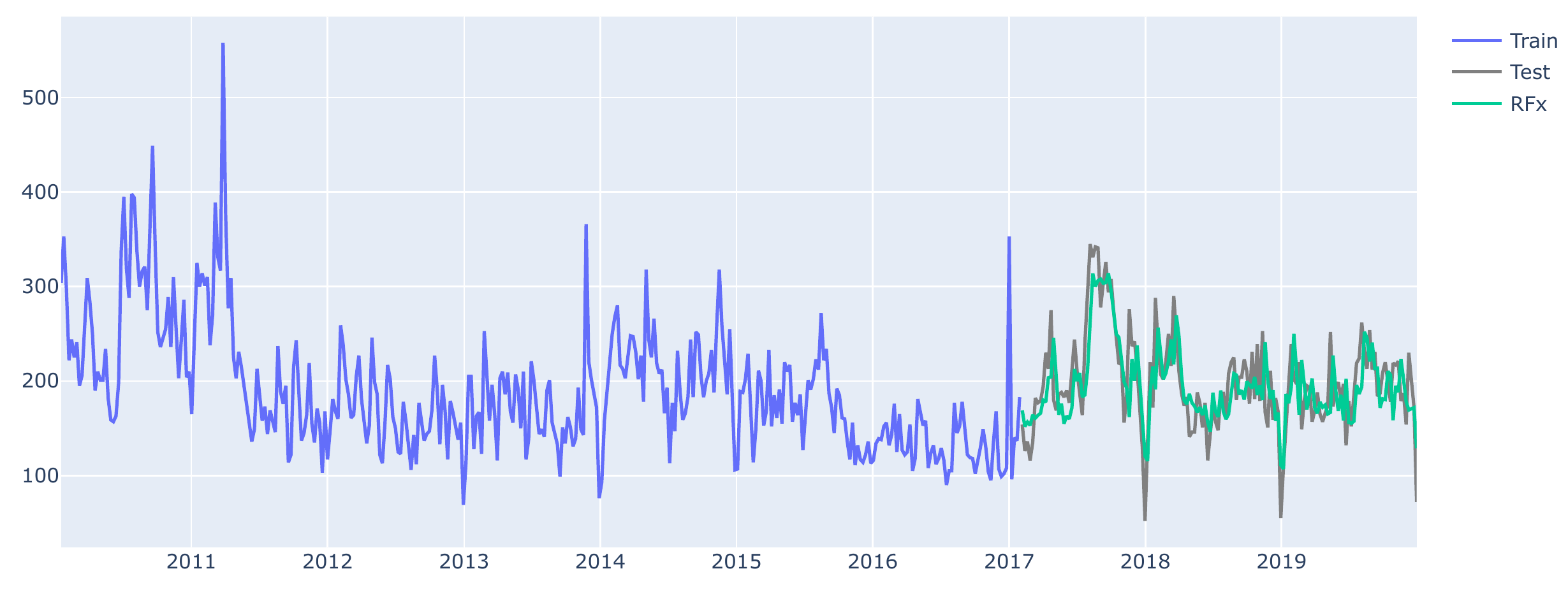}
    \caption{A sample output of the bug number prediction for the last 3 years}
    \label{fig:rf}
\end{figure*}

Table~\ref{tab:result_all} shows the performance of each method. Random Forest with exogenous variables (RFx) has the best performance in terms of RMSE, error percentage and standard deviation of the errors. Surprisingly, the simple exponential smoothing generates reliable predictions in terms of MAE and $R^2$. We observe that adding exogenous variables does not necessarily augment the performance of the algorithms, including ARIMA and LSTM. One of the reasons could be the sensitivity of those algorithms to the new features. They cannot differentiate between the time series and exogenous variables; thus, the effect of the values of previous time steps will be eclipsed by the newly defined variables. On the other hand, in the software engineering domain, the Random Forest is proved to be robust, highly accurate, and resilient to noisy data~\cite{Jiang2008,kamei2016}. Therefore, it can deal with new features that might be unimportant or have a negligible effect on the output. RFx has a better performance than a simple RF Regressor as more exogenous features will reduce its bias. Figure~\ref{fig:rf} shows the general overview of the training-test split and predicted time series using Random Forest Regressor with covariates.

\setlength{\tabcolsep}{8pt}
\renewcommand{\arraystretch}{1.2}
\begin{table}[!ht]
\centering
\caption{The result of different time series prediction models in terms of RMSE, $R^2$, Error percentage, MAE, and Std. The best value for each metric is bold-faced. \label{tab:result_all}}%
\resizebox{0.85\linewidth}{!}{
\begin{tabular}{lrrrrr}
\toprule
\textbf{Method} & \textbf{RMSE} & \textbf{R-squared} & \textbf{Error (\%)} & \textbf{MAE}& \textbf{Std} \\
\midrule
\textbf{EXP} & 39.36 & \textbf{0.114} & 0.178 & \textbf{22.09} & 39.36 \\\hline
\textbf{WMA} & 39.26 & 0.270 & 0.181 & 25.17 & 39.25 \\\hline
\textbf{ARIMA} & 37.46 & -0.131 & 0.170 & 22.62 & 37.43 \\\hline
\textbf{ARIMAx} & 42.97 & -0.448 & 0.199 & 29.24 & 42.92 \\\hline
\textbf{LSTM} & 39.86 & 0.127 & 0.185 & 25.07 & 39.78\\\hline
\textbf{LSTMx} & 42.29 & 0.383 & 0.194 & 25.81 & 42.29 \\\hline
\textbf{RF} & 40.18 & 0.167 & 0.171 & 23.30 & 39.87 \\\hline
\textbf{RFx} & \textbf{36.06} & 0.179 & \textbf{0.160} & 24.39 & \textbf{35.88} \\
\midrule
\textbf{Base} & 41.04 & 0.346 & 0.183 & 27.00 & 41.03\\
\bottomrule
\end{tabular}
}
\end{table}

\begin{figure*}[ht!]
    \centering
    \includegraphics[width=1.01\textwidth]{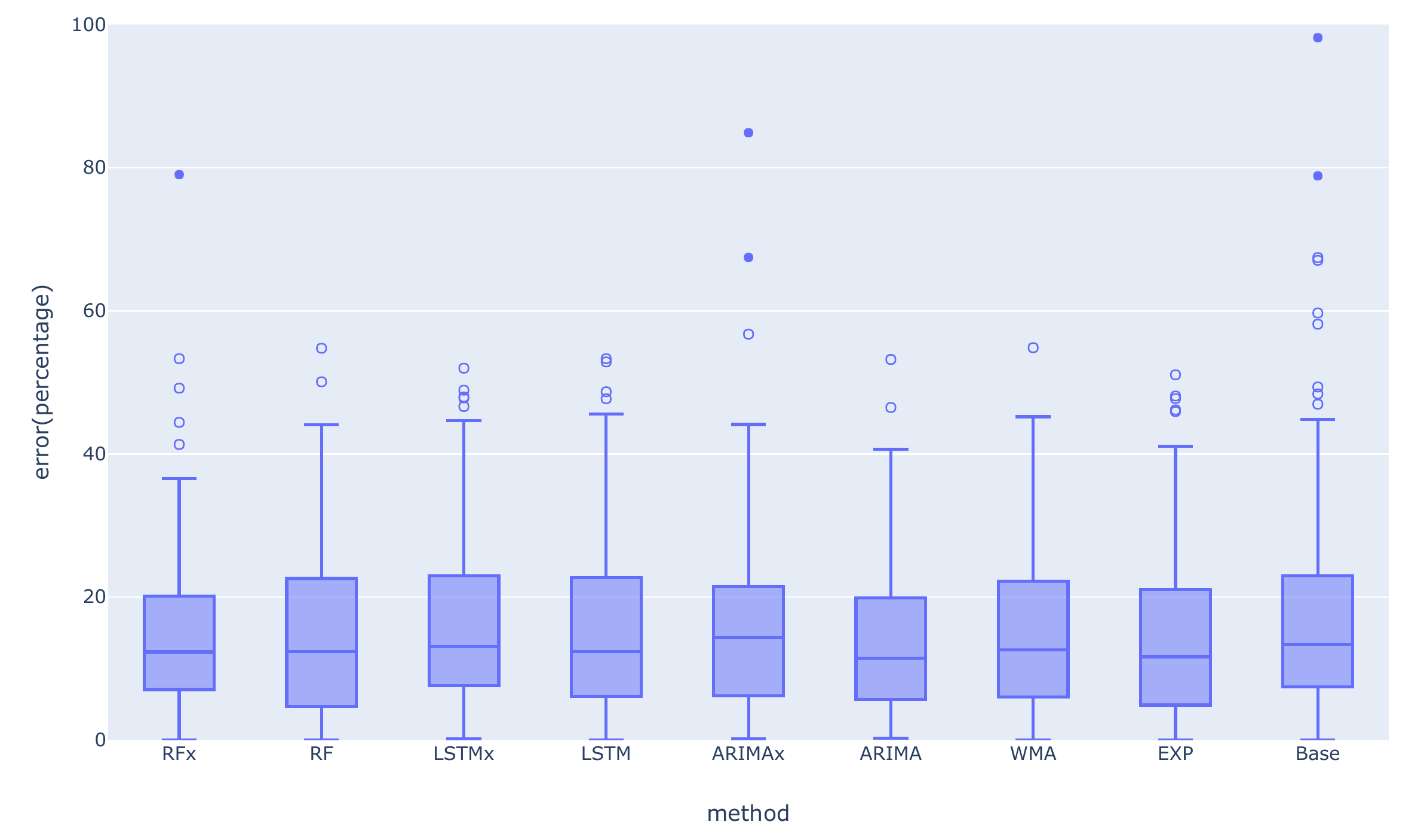}
    \caption{Prediction performance of different algorithms for one-step prediction (based on error percentage)}
    \label{fig:time_series_boxplot}
\end{figure*}

The answer to the RQ1 is that due to random fluctuation in the number of bugs introduced to the Mozilla project, the performance of the proposed models remains almost the same. However, Random Forest with exogenous features outperforms the others in most cases with the least variance (see Figure~\ref{fig:time_series_boxplot}).

\begin{figure*}[ht!]
    \centering
    \includegraphics[width=1.01\textwidth]{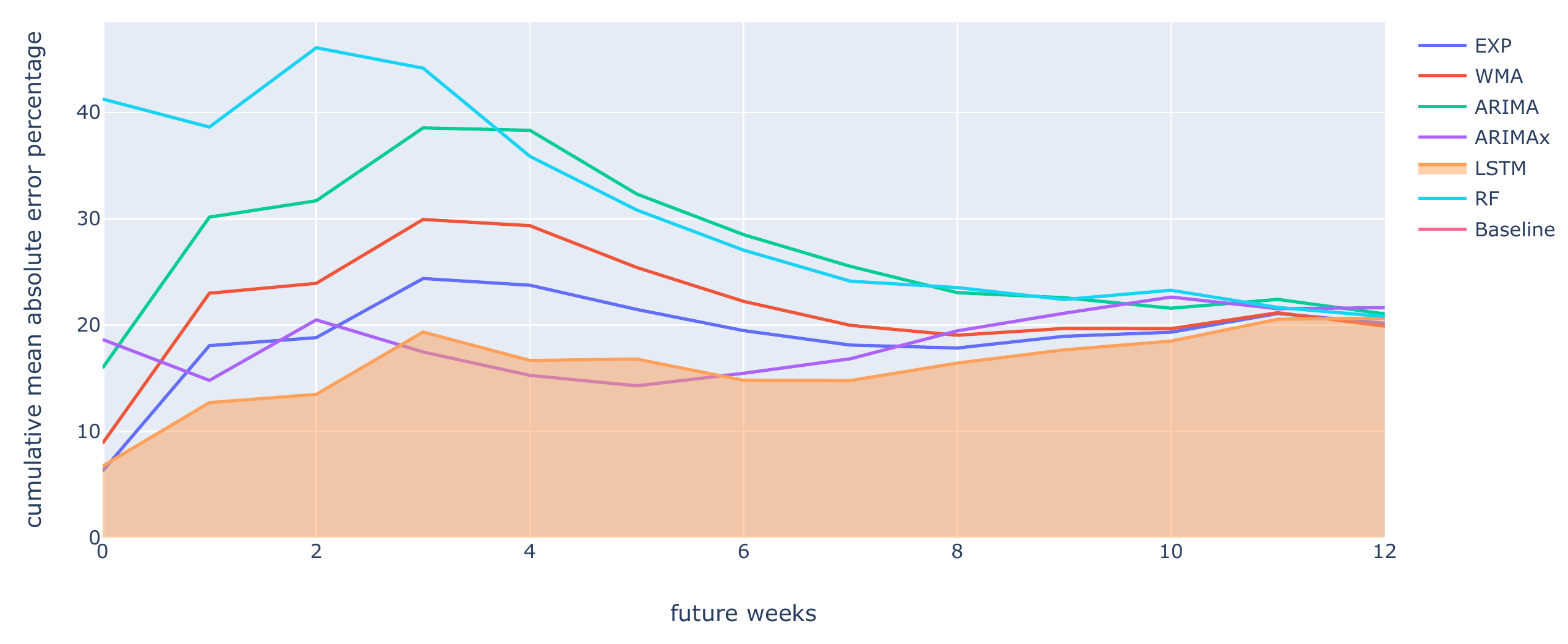}
    \caption{3-month prediction performance of the models, reported as cumulative mean absolute error percentage}
    \label{fig:future_bug}
\end{figure*}

In RQ1, we investigated the prediction performance of the models for the next step (next week) whereas in the RQ2 we want to see the effect of long-term prediction. As for the long-term prediction, the error of the models is accumulated, the future predictions would not be as accurate as the closer ones. Here, we analyse the performance of the models for a 3-month bug number prediction. Figure~\ref{fig:future_bug} shows that LSTM with the capability of having both Long and Short Term Memory performs at the same level in long-run without compromising accuracy. This merit makes it the most suitable model for long-term prediction.

\section{Threats to validity}\label{sec:threats}
In this section, we disclose threats to the validity of our empirical study.

\subsection{Construct Validity}
We estimated the performance of the models using a train-test rolling approach. However, the rolling approach can be used when we have enough history of data. As cross-validation can not be directly applied in time series prediction, the rolling strategy maintains the chronological order of the series and increases its reliability. Other experiment designs, such as nested rolling Cross-validation\cite{Varma2006}, may yield different results. 

Although we study four different exogenous variables, there are likely more external features (e.g., major release or social media rumours about a defect in a project) that may impact the trends in the dataset. We plan to expand our exogenous variables to include additional external factors since some previous works that the correct implementation of exogenous terms will improve the performance of time series prediction~\cite{Xiao2019}.

\subsection{Internal Validity}
The tracking dataset data is extracted from the Bugzilla using the REST API\footnote{https://wiki.mozilla.org/Bugzilla:REST\_API}. While extracting the dataset, we consider all records between Jan 2010 and Dec 2019 to have the most recent number of reported bugs. However, some issue records might be removed from the repository or imposed restricted access to normal users. We ensure to extract all publicly available bugs; however, there might be some other defects that are only visible to developers. 

\subsection{External Validity}
We only consider the biggest project in Bugzilla, and hence, our result may not be generalizable to all software systems. However, the dataset is a large, long-lived system that alleviates the likelihood of bias in our report. Nonetheless, replication of our study using additional systems may prove fruitful~\footnote{https://github.com/HadiJahanshahi/Bug-Number-Prediction}. 

We used 8 different models to evaluate the feasibility of bug number prediction across the projects whereas using other forecasting techniques may boost the prediction performance.  

\section{Conclusion and Future Work}\label{sec:conclusion}
In this paper, we study the effectiveness of a time series prediction methods in predicting the number of bugs reported to a software repository. Our main aim is to help practitioners anticipate an abnormal number of bugs introduced in a specific time and be well prepared by planning ahead. Predicting the number of bugs discovered in the system may also provide insights for developers and managers about the trends of discovered defects and therefore the quality of software. Time series analyses have two-fold outcomes: first, to forecast the number of future defects that may occur in the system; second, to identify the trends and abnormality in the system.

Through an empirical study on the number of bugs introduced to the Mozilla project, we made the following observations:
\begin{itemize}
    \item The number of bugs introduced to the aforementioned system does not have a unit root and therefore is stationary. Furthermore, no specific trend has been observed in the dataset. 
    \item Using eight different forecasting methods, we conclude that there are some improvements in the prediction performance compared to the baseline. Considering five different metrics, Random Forest with exogenous variables exceeds other methods. Nevertheless, we expected to observe a significantly better performance while all models have almost the same error distribution.
    \item Using a neural network, especially LSTM, significantly improve the long-term prediction. LSTM is not affected by the residual error of the prediction when applied on the long-term test dataset whereas other methods construct their prediction on the error of the last period; hence they are unable to challenge the robustness of LSTM. 
\end{itemize}

\bibliographystyle{splncs04}
\bibliography{main}

\end{document}